\pdfoutput=1
\documentclass{PoS}
\graphicspath{{include//}}
\usepackage{caption}
\usepackage{subcaption}						
\usepackage{placeins}   
\usepackage{amsmath} 
\newcommand*{\doi}[1]{doi: \href{http://dx.doi.org/#1}{#1}}

\title{Studies towards an understanding of global array pointing for the Cherenkov Telescope Array}

\ShortTitle{CTA pointing studies}

\author{Stefan Eschbach$^a$, \speaker{Alexander Ziegler} $^a$, Christopher van Eldik$^a$, Johannes Veh$^a$, David Berge$^{b}$ and Markus Gaug$^{c}$ for the CTA Consortium\footnote{Full consortium author list at http://cta-observatory.org}
\\
        \llap{$^a$}Erlangen Center for Astroparticle Physics (ECAP), Universit\"at Erlangen-N\"urnberg, Germany\\
        \llap{$^b$}Anton Pannekoek Instituut voor Sterrenkunde,
        Universiteit van Amsterdam, Netherlands\\
        \llap{$^c$}Unitat de F\'isica de les Radiacions, Departament de F\'isica, and CERES-IEEC, Universitat Aut\`onoma de Barcelona, E-08193 Bellaterra, Spain\\
        E-mail: \email{stefan.eschbach@fau.de}}	
	
\abstract{Current arrays of Imaging Atmospheric Cherenkov Telescopes (IACTs)
	routinely achieve an astrometric point-source location accuracy of 20-30
	seconds of arc (given large photon statistics), which is well below the
	angular resolution obtained for individual photons. The location accuracy
	is mainly limited by systematic uncertainties due to possible deformations
	of the telescopes' structures, causing a mis-orientation between the
	nominal optical axes of the telescopes and their actual pointing
	directions. Usually, only a subset of telescopes is triggered for a given
	gamma-ray event, and their mis-orientations enter the gamma-ray direction
	reconstruction in a combined manner. Hence, in general, the average
	location accuracy for a set of gamma-ray observations depends in a complex
	way on the individual telescope accuracy and the geometry of the array,
	but also on the observing conditions, the energy spectrum of the source
	and the chosen event selection cuts.
	
	For the proposed Cherenkov Telescope Array (CTA), a post-calibration
	point-source location accuracy of 3 seconds of arc is aimed for under
	favorable observing conditions and for gamma-ray energies exceeding 100
	GeV. In this contribution, results of first studies on the location
	accuracy are presented. These studies are based on a toy Monte Carlo
	simulation of a typical CTA-South array layout, taking into account the
	expected trigger rates of the different CTA telescope types and the
	gamma-ray spectrum of the simulated source. With this simulation code it
	is possible to study the location accuracy as a function of arbitrary
	telescope mis-orientations and for typical observing patterns on the sky.
	Results are presented for various scenarios, including one for which all
	individual telescopes are randomly mis-oriented within their specified
	limits. The study provides solid lower limits for the expected source
	location accuracy of CTA, and can be easily extended to include various
	other important effects like atmospheric refraction or partial cloud
	coverage.
	}

\FullConference{The 34th International Cosmic Ray Conference,\\
		30 July- 6 August, 2015\\
		The Hague, The Netherlands}

\begin{document}

\section{Introduction}
To resolve the exact position of gamma-ray sources on the sky, a high localisation accuracy of the observing instrument is key.
The typical angular resolution of an Imaging Atmospheric Cherenkov Telescope (IACT) is $\theta\sim 0.1^{\circ}$ for a single gamma-ray photon \cite{Actis2011}, limited by the finite size of camera pixels, instrumental coverage of the Cherenkov light pool on ground, and, ultimately, fluctuations in the shower development. The statistical source location accuracy, however, which is given by the centroid of the reconstructed directions of all gamma rays detected from the source, is significantly better: it is roughly given by $\theta/\sqrt{n}$, with $n$ being the number of detected gamma-ray events. Sources of high intensity can thus be located with a statistical accuracy of well below 10".

Another possible source of localisation uncertainty is caused by imprecise pointings of the individual telescopes of the IACT array. Such telescope pointing errors are systematic errors that emerge from mechanical deformations of the telescopes or because of small offsets during the tracking of the gamma-ray source. Especially for high-intensity gamma-ray sources, where the statistical error on the source location is small, a good understanding of the pointing systematics is crucial for an accurate determination of the source position. Here, we assess the expected source localisation uncertainty of CTA by means of toy Monte-Carlo simulations of gamma-ray showers detected by telescopes with artificially-introduced pointing offsets.

In ground-based gamma-ray astronomy, it is common to perform detailed Monte-Carlo simulations of the gamma-ray detection mechanism, from the development of electromagnetic showers in the atmosphere, via the propagation of Cherenkov light, to the detection of the Cherenkov photons in the cameras of the telescopes. In principle, the impact of telescope pointing errors on the reconstructed shower directions can be investigated with these simulations. Their drawback, however, is that they are time consuming and need lots of computing power to generate a gamma-ray sample large enough such that statistical uncertainties on source localisation become negligible.
To quantify systematic errors caused by mis-pointing isolated from other effects (e.g.\ smearing due to limited angular resolution) it is therefore reasonable to reduce complexity to a lower level. 
The simulation presented here pursues an approach where a gamma-ray shower is reduced mainly to the major axis of the corresponding Hillas ellipse and its primary energy. With this simplification it is possible to simulate thousands of showers with minimal computing time. For each shower, lookup tables based on full Monte-Carlo simulations of the array determine which telescopes are able to detect the shower, and telescopes are triggered randomly based on probability distributions. The shower arrival direction is then reconstructed geometrically using the information of all telescopes which participated in the detection. Telescope pointing errors can be artificially applied to the telescopes to study their impact on the reconstructed arrival direction. 
The simulation presented here is not restricted to simulating at fixed zenith and azimuth angles. Instead, the gamma-ray source position is given in the RA/Dec system and tracked during the observation, while moving across the sky. As a result, rotations of the field of view and their influence on the reconstruction in the presence of pointing errors are also simulated. Wobble offsets are taken into account.

\section{Simulation}
The code for the simulation presented in this proceeding is using ROOT \cite{root}, an object-oriented framework to analyse event-based data, based on C++. In the following, it will be described in detail how the showers and telescope responses are simulated, and how the detection and shower direction reconstruction is implemented.
\subsection{Simulated showers}
As stated before, gamma-ray showers in this simulation are solely represented by primary energy and orientation of the shower axis in the atmosphere. The first point on the shower axis is given by the RA/Dec position of the source. By choosing a start and end time for the simulated observation, the position of the source in the Alt/Az system can be calculated at each time step of the observation while the source is moving across the sky. The second point on the shower axis is the shower core position, which is defined as the intersection point of the shower axis with the ground. The core position is chosen randomly, uniformly distributed in a circular area perpendicular to the telescopes' pointing direction.

The energy $E$ for each shower is chosen randomly using
\begin{equation*}
E= E_0 \cdot x^{1/\left( 1-\Gamma \right) },
\end{equation*}
where $x$ is a random number uniformly distributed between 0 and 1, resulting in a power-law spectral energy distribution $\text{d}N_{\gamma}/\text{d}E$ with spectral index $\Gamma$. No showers are simulated for energies $E < E_0=5$\,GeV, since CTA is not sensitive for such low energies. Note that the spectral index of the gamma-ray spectrum has a strong impact on the resulting source localisation uncertainty, because it determines the average number of telescopes that participate in the detection of an event.

Lastly, a reduced penetration depth $D=x/x_0$, which is the path length that the primary photon travelled through the atmosphere before the very first interaction, is chosen randomly for each incident gamma-ray after an exponential distribution. With the penetration depth and the energy, the height of the shower maximum, conventionally expressed in units of g~$\text{cm}^{-2}$, can be calculated to be
\begin{equation*} 
X_{max}= \left( D +  \ln\left(\frac{E}{E_c}\right)\right) \cdot x_0, 
\end{equation*} 
with the radiation length $x_0=36.62$~g~$\text{cm}^{-2}$ and the critical energy for pair creation $E_c=85$\,MeV.\\ 
$X_{max}$ can easily be converted to height over ground level by applying the barometric height formula with scale height $h_s=6532.39$\,m.\\

\subsection{Simulated telescopes}
\begin{figure}
	\centering
	\includegraphics[width=0.4\textwidth]{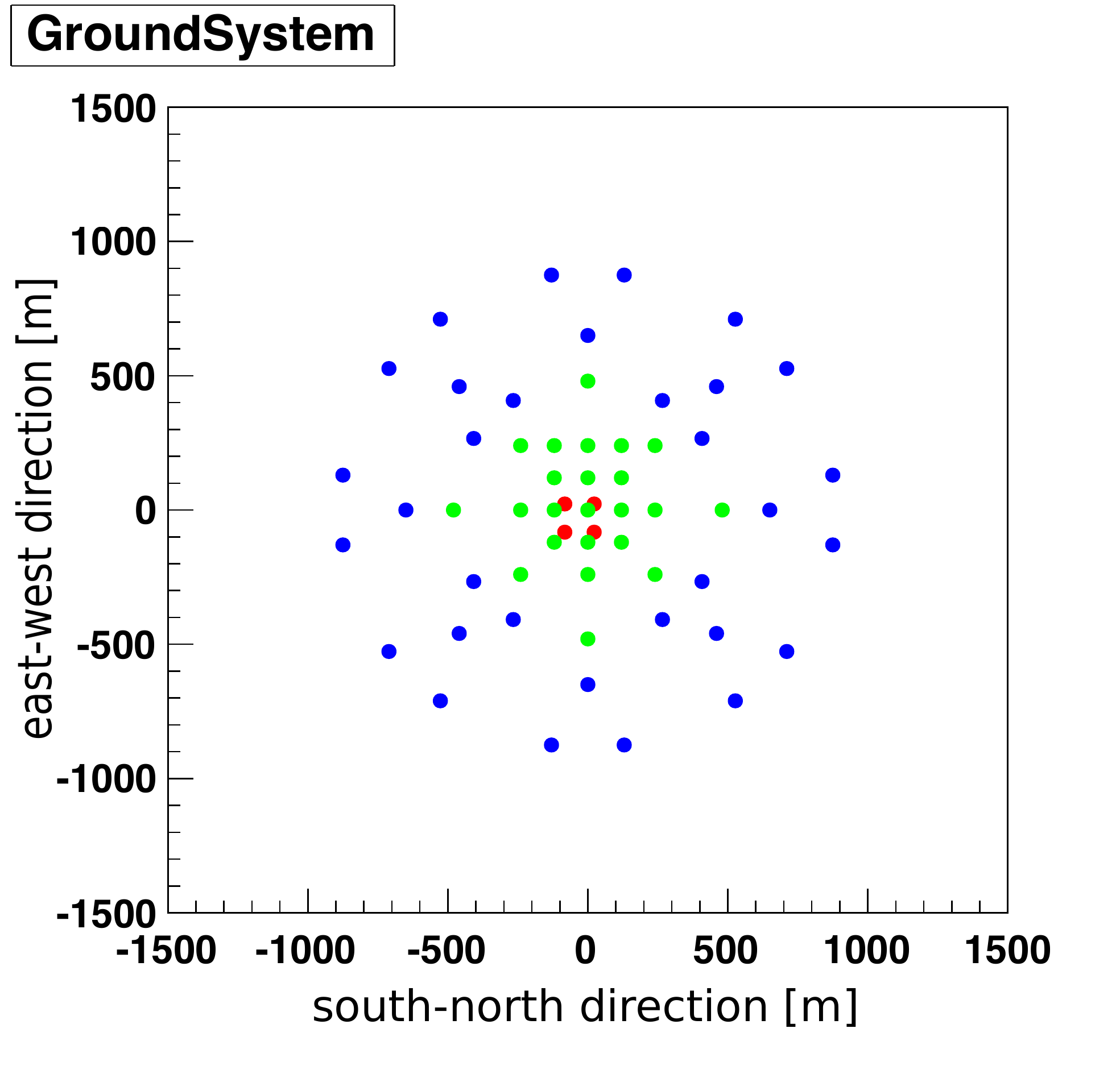}	
	\caption{Array E \cite{Bernloehr2013}, a typical CTA-South array layout, seen from above. Axes give the distance to the array center in meters. The colors represent the telescope types: 4 LSTs (red), 23 MSTs (green) and 32 SSTs (blue).}
	\label{ArrayE}
\end{figure}

In the simulation, each telescope is defined by a unique ID, its type (Large Sized Telescope (LST), Medium Sized Telescope (MST) or Small Sized Telescope (SST)), its position on the ground and possibly individual pointing errors. The parameters for a whole array of telescopes are stored in a text file, which can easily be changed to simulate different pointing errors or even other array layouts. The layout used in order to produce the results presented in this proceeding is 'Array E' \cite{Bernloehr2013}, an early proposed layout for CTA South, which is shown in Figure~\ref{ArrayE}. For the results presented here, the telescopes are pointed directly to the source position in RA/Dec coordinates (modulo artificial pointing errors) and are tracking the source during its movement. The relative impact of different pointing errors on the reconstructed source position may therefore change over time. 

\subsection{Detection with lookup tables}
\begin{figure}[h]
	\begin{center}
		\includegraphics[width=0.4\textwidth]{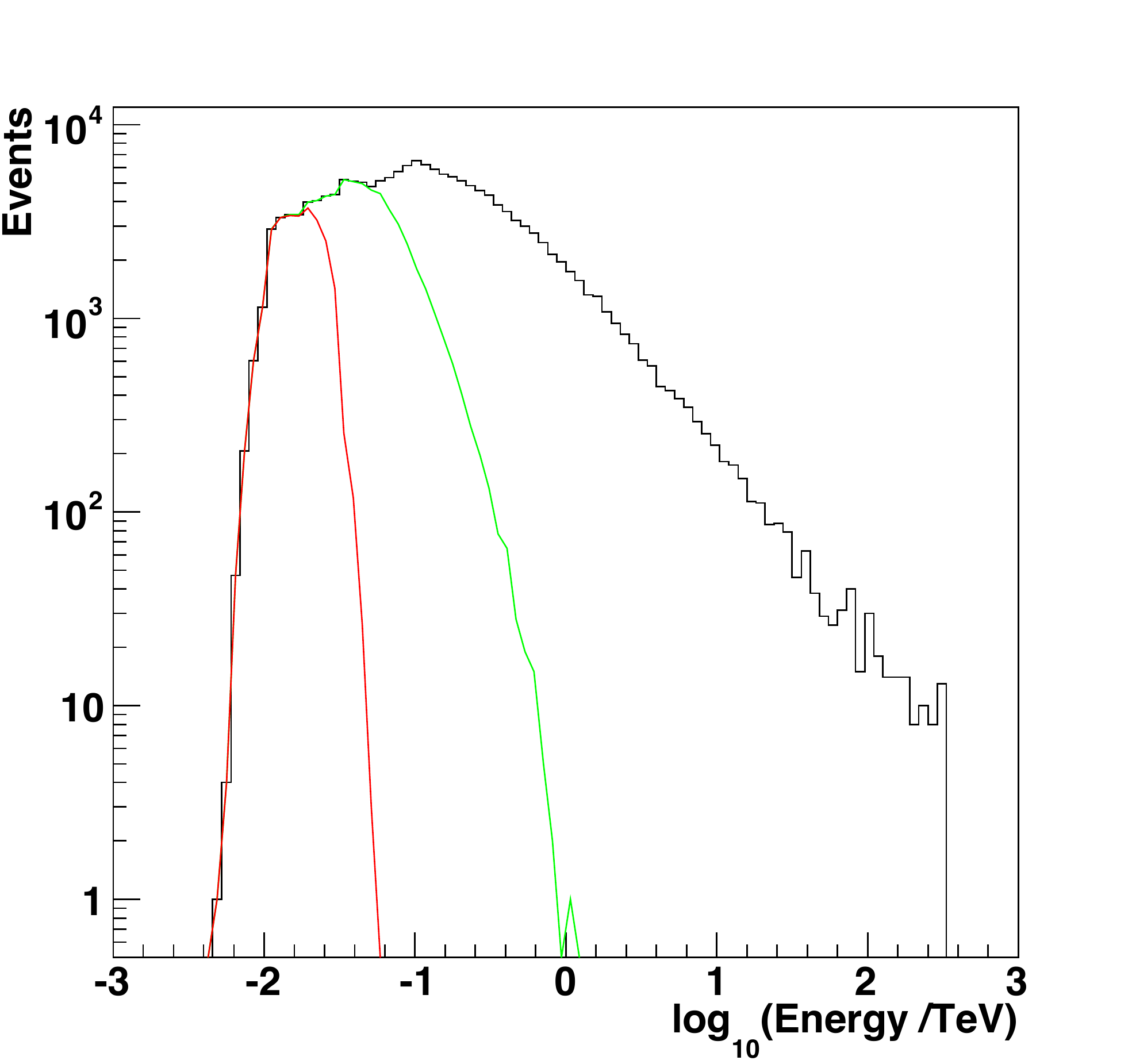}
		\caption{Distribution of reconstructed showers as function of gamma-ray energy for 4\,000\,000 simulated showers, double logarithmic scale. Black: Distribution of energies of all detected showers. Red: Showers detected exclusively by LSTs. Green: Showers detected by LSTs and/or MSTs (no SSTs). At high energies, mostly all showers are detected and the shape of the distribution follows the simulated source spectrum.}
		\label{perfectenergy}
	\end{center}
\end{figure}

For each simulated shower and each telescope, it is first checked whether the shower is in the telescope's field of view. If this is the case, the number of photoelectrons (image amplitude) the shower creates in the telescope camera is determined via lookup tables. The image amplitude depends on the shower energy $E$, the shower penetration depth $X_{max}$, the distance between its core position and the telescope position, the zenith angle and the telescope type. If the image amplitude exceeds a certain detection threshold, 
the shower is rated as being detected by this telescope. The amplitude threshold cut chosen was 50 photo electrons, which is a compromise between detecting a sufficient number of showers and guaranteeing safe threshold energies for the different telescope types (see Figure~\ref{perfectenergy}). At least two telescopes have to detect a certain shower for the event to pass the selection and being counted as detected by the array (stereo reconstruction).\\ 
The lookup tables used are produced with an analytical model, based on full Monte Carlo simulations of the telescopes \cite{Naurois2009}. All studies presented here were performed at a zenith angle between 6.7$^{\circ}$ and 8.8$^{\circ}$.

\subsection{Direction reconstruction}
To reconstruct the direction of a shower, the data of all telescopes which contributed to its detection are used. The shower axis is transformed into the camera plane of each telescope, taking into account the chosen pointing offsets. The shower axes are then intersected pair-wise in a common camera coordinate system, and a common intersection point $\vec{S}$ is calculated by weighted averaging: 
\begin{equation*}
\vec{S}=\frac{\sum_{k>j} \vec{S}_{j,k} \cdot \sin \alpha_{j,k} \cdot \frac{1}{\frac{1}{A_j}+\frac{1}{A_k}}}{\sum_{k>j} \sin \alpha_{j,k} \cdot \frac{1}{\frac{1}{A_j}+\frac{1}{A_k}}},
\end{equation*} 
where $\vec{S}_{j,k}$ is the intersection point of the shower axes recorded by telescopes $j$ and $k$, $\alpha_{j,k}$ is the respective intersection angle, and $A_{j}$ respectively $A_{k}$ the image amplitudes seen by the telescopes.
The common intersection point, which defines the reconstructed source position in camera coordinates, is then transformed to the RA/Dec system to get the reconstructed source position in sky coordinates. Without pointing offsets introduced in the simulation, the reconstructed and true source position are identical.

\section{Results}
\FloatBarrier

\begin{figure}[htbp]
	\centering
	\includegraphics[width=0.48\textwidth]{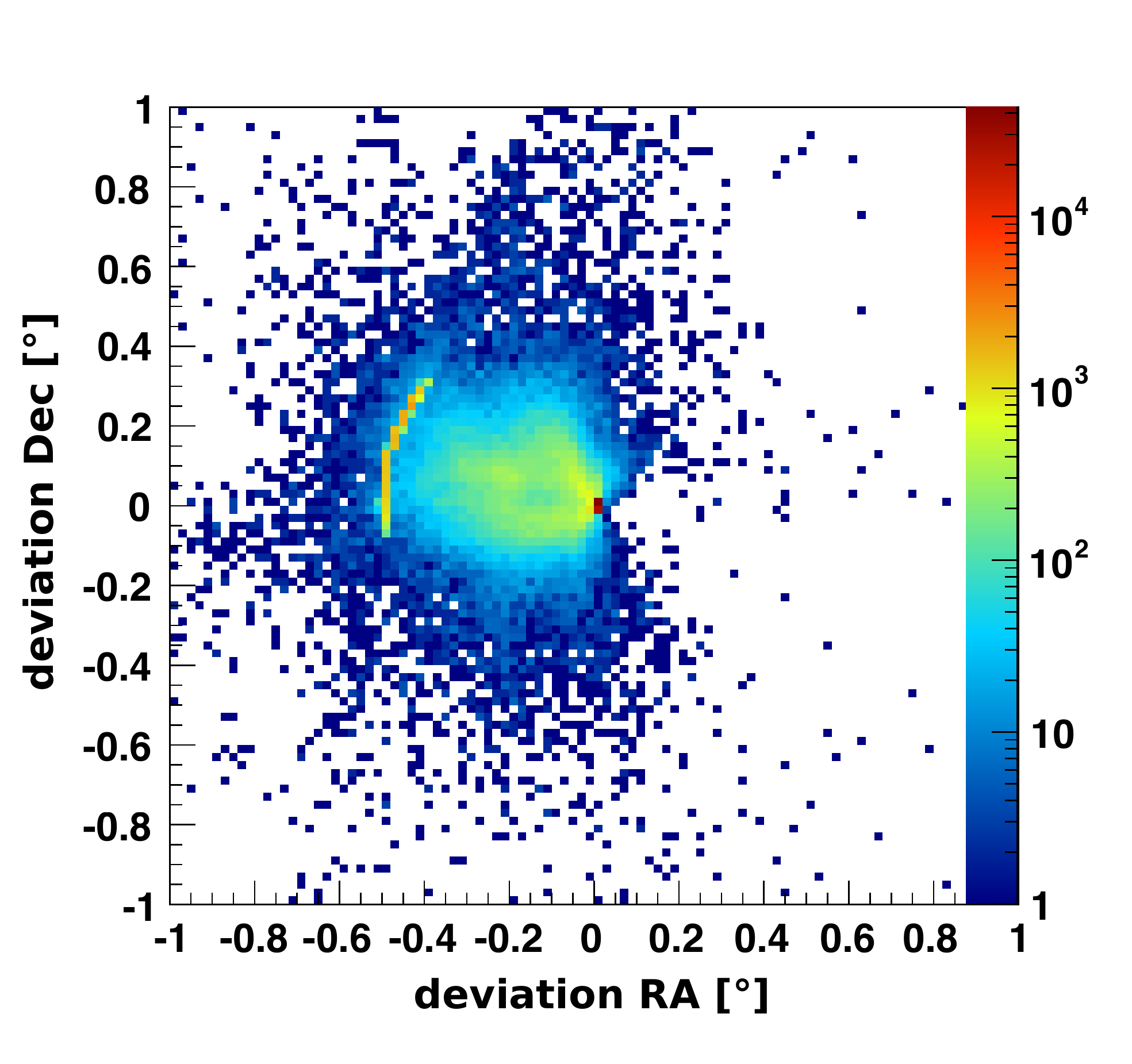}
	\caption{Distribution of reconstructed shower directions. The color scale is logarithmic and gives the number of showers detected in a certain bin. A pointing offset of -0.5$^{\circ}$ in the altitude has been applied to only the LSTs. Many showers are reconstructed at the true source position, when no LST takes part in their detection. For detections where different telescope types (including the LSTs) were involved, a shift to negative RA values can be seen. At 0.5$^{\circ}$ distance to the true source position, an arc-like substructure is visible, which is due to events solely detected by the LSTs. Here one can see the rotation of the field of view due to mis-tracking of the source because of the applied pointing offsets. The mean reconstructed source position has a deviation of $\alpha=(-0.1444 \pm 0.0006)^{\circ}$ in RA and $\delta=(0.0480 \pm 0.0005)^{\circ}$ in Dec from the true source position.}
	\label{lstzoom}
\end{figure}

With the simulation described above, a study with numerous combinations of pointing offsets has been performed. A small selection of results is presented here. The simulations have been performed with a spectral source index of $\Gamma=2$. To get a first impression of the performance of the simulation, only the four LST telescopes were equipped with the same pointing offset of 0.5$^{\circ}$ in negative altitude direction, simulating a sagging of camera masts because of their own weight. Note that this offset is by far larger than expected for any real CTA telescope, but makes the impact on the source position reconstruction easily visible. In Figure~\ref{lstzoom}, the reconstructed directions of $10^6$ such simulated showers are shown. There are three regimes visible: Showers detected without the participation of LSTs are reconstructed correctly at the true source position, showers detected by a mixture of telescope types (including LSTs) get reconstructed wrongly, with a general shift to negative RA coordinates (corresponding to the applied pointing offsets for the chosen source and array location). Showers detected solely by LSTs are visible as an arc with a deviation of 0.5$^{\circ}$ from the true source position. The arc-like form is due to the rotation of the field of view because the source was not tracked correctly any more.\\

To quantify the influences of real pointing errors, where not necessarily all telescopes suffer from the same offset, random offsets have been applied, generated from a Gaussian distribution centered at zero. The width of the distribution was chosen differently for the different telescope types: For MSTs and SSTs, it was set to $\sigma=7''$, for LSTs it was set twice as wide to $2\sigma=14''$. These numbers match the telescope-wise values for the post-calibration pointing precision aimed for in the design of CTA.

\begin{figure}[htbp]
	\begin{subfigure}[t]{0.45\textwidth}
		\centering
		\includegraphics[width=\textwidth]{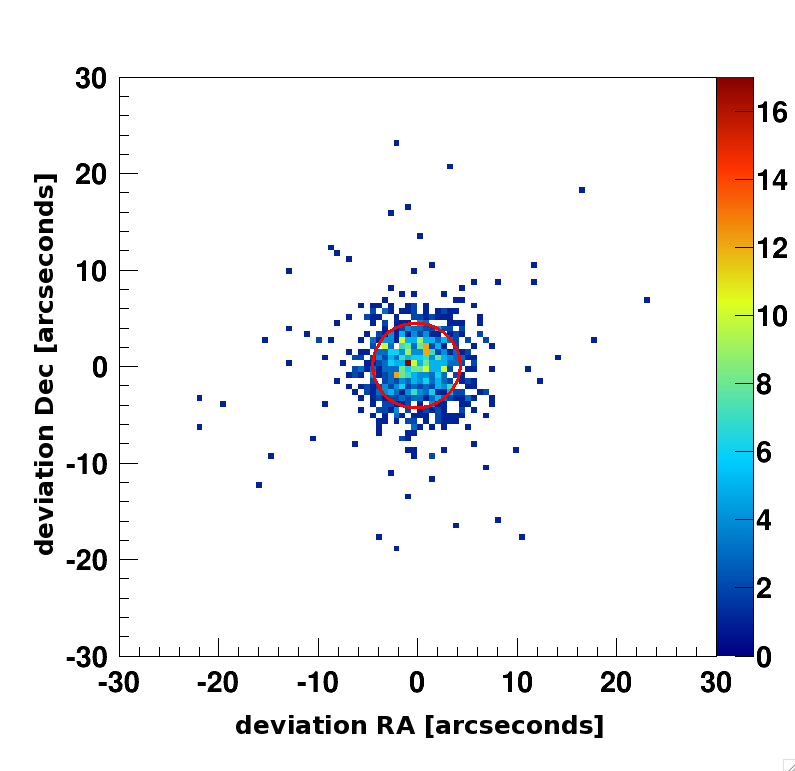}
		\caption{Deviations from the true position of 1000 reconstructed source positions. Each reconstructed source position is the mean of thousands of reconstructed shower directions with a certain set of randomly chosen pointing offsets. The red circle contains 68\% of all reconstructed source positions and has a radius of $r_{68}=4.4"$.}
		\label{random7index2radec}
	\end{subfigure}
	\hfill
	\begin{subfigure}[t]{0.45\textwidth}
		\centering
		\includegraphics[width=\textwidth]{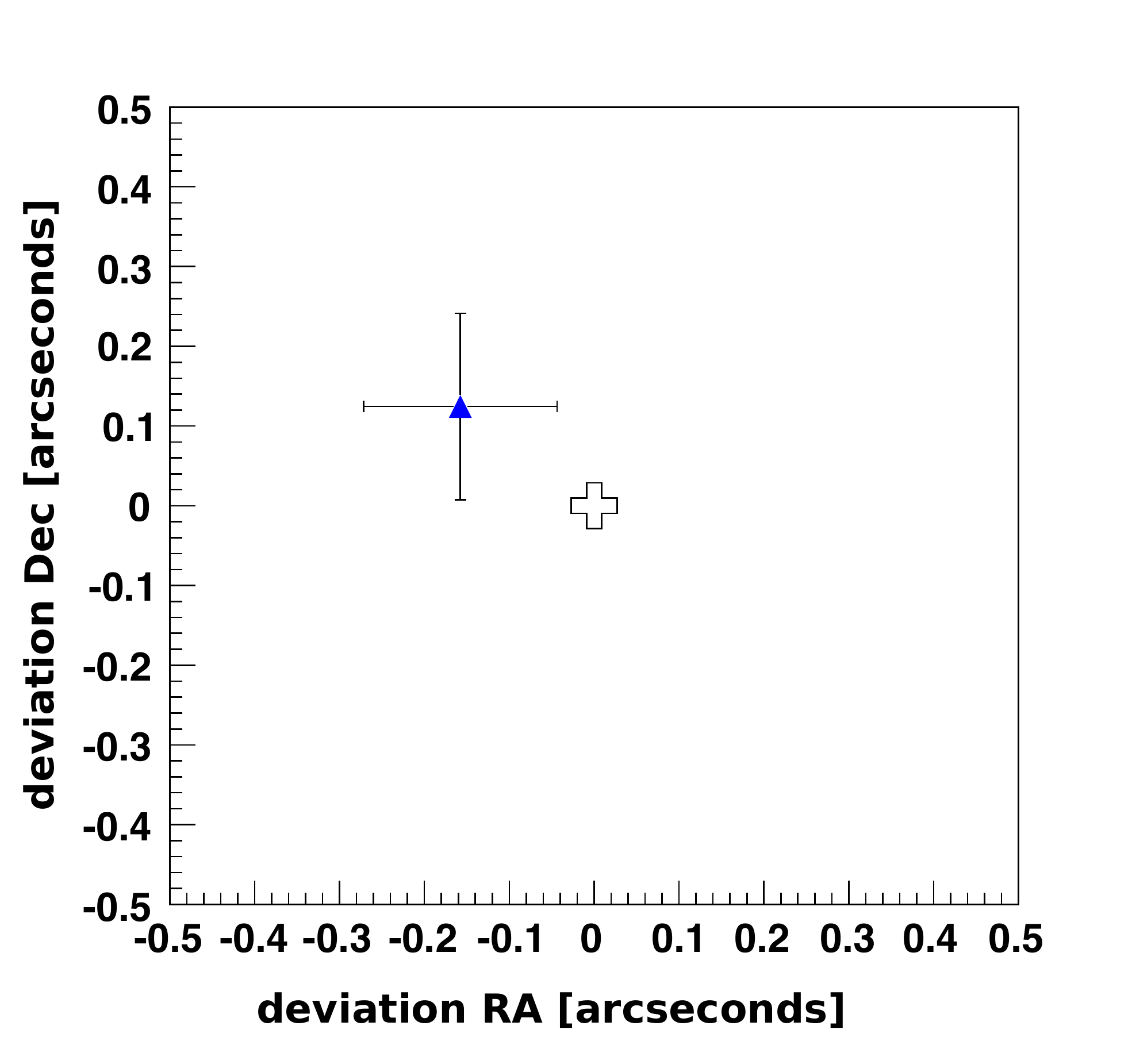}
		\caption{Mean reconstructed source position (blue triangle), which is the centroid of the distribution shown in Figure~\ref{random7index2radec}. The deviation of the true source position (black cross) is $(-0.158 \pm 0.114)"$ in RA and $(0.125 \pm 0.117)"$ in Dec. The linear distance is $r_{\text{shift}}=(0.201 \pm 0.115)"$.}
		\label{random7index2shift}
	\end{subfigure}
	\caption{Results of $10^3$ simulation runs with different random offsets.}
\end{figure}

With a fixed set of random offsets, a single simulation run with $10^4$ simulated showers was executed, and the reconstructed source position for this run was calculated as the mean of all reconstructed shower directions. After that, new random offsets were chosen and another run was started. In Figure~\ref{random7index2radec}, the distribution of reconstructed source positions of $10^3$ such simulation runs is shown.
The typical width of the distribution, quantified in terms of the 68\% containment radius, is a measure for the pointing-error induced systematics on the position reconstruction of a single observation.
For the case shown here, with random pointing offsets in the range of $14''$ (LSTs) and $7''$ (MSTs and SSTs), respectively, the systematic source localisation error is about $r_{68}=4.4"$. This is less than the post-calibration point-source localisation accuracy CTA is aiming for, namely $10"$ per axis for energies below 100 GeV.



\begin{figure}[htbp]
	\begin{centering}
		\includegraphics[width=0.40\textwidth]{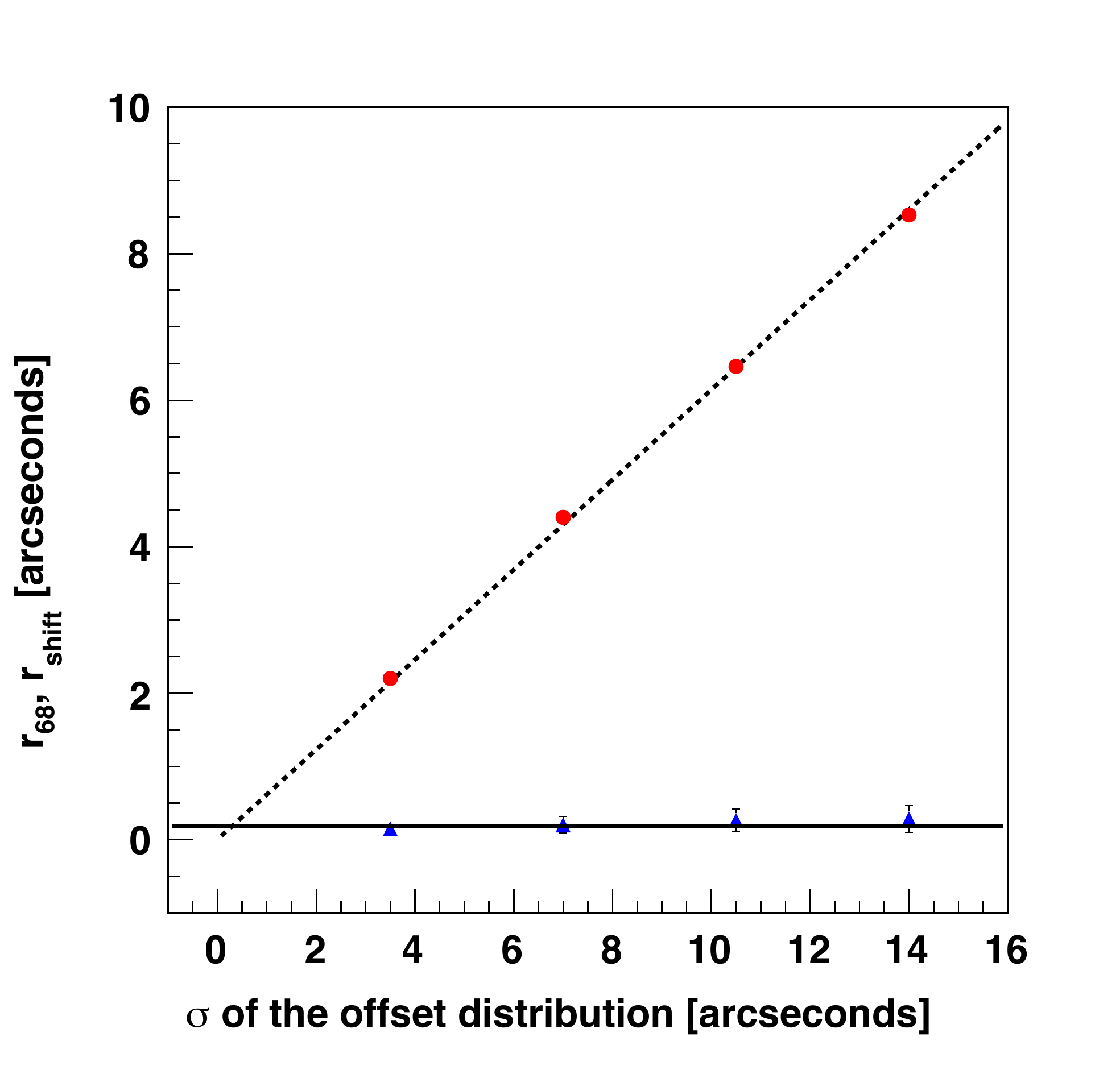}
		\caption{The parameters describing the distribution of reconstructed source positions, $r_{68}$ and $r_{\text{shift}}$, as a function of the width $\sigma=\sigma_{SST}=\sigma_{MST}=\frac{1}{2}\sigma_{LST}$ of the Gaussian pointing offset distribution.\\
		Red circles mark $r_{68}$, which is fitted with a straight line through the origin with a slope of $m=0.61$ (dashed line). Blue triangles mark the development of $r_{\text{shift}}$, which is fitted with a constant line at $r_{\text{shift}}=(0.19\pm0.06)"$ (solid line).}
		\label{randomsigma}
	\end{centering}
\end{figure}

In Figure~\ref{randomsigma}, the dependency of $r_{68}$ on the width of the offset distribution is shown. It demonstrates that even if telescope pointing errors are doubled, results are still below CTA design goals. However, this simulation provides only a lower limit estimate, not yet factoring in additional effects worsening the location accuracy, like e.g. atmospheric refraction or the impact of the Earth's magnetic field. 

Note that, when averaging over many observations, the mean reconstructed source position should be compatible with the true position, if a symmetric distribution of pointing errors is assumed (cf.\ Figs.~\ref{random7index2shift} and \ref{randomsigma}). However, this is not necessarily the case for real observations, since individual telescope pointing errors are expected to be partially correlated (e.g. due to bending of the telescopes into the same direction under the influence of gravity). 

\section{Conclusion}
In this proceeding, a simulation used to quantify the influence of telescope pointing errors on the direction reconstruction of CTA is presented. Although it simplifies the shower development in the atmosphere to a straight line describing the main shower axis, it can be used to set solid lower limits on the expected source location accuracy of CTA for given post-calibration pointing accuracies of the telescopes. The method can easily be adjusted for arrays other than CTA and can be extended to take into account various other effects like atmospheric refraction or to study other pointing systematics like mechanical rotations of the cameras.

It is shown that, assuming randomly Gaussian distributed pointing offsets of $7"$ (SSTs, MSTs) and $14"$ (LSTs), the lower limit for the source location accuracy is approximately $4.4"$. Due to the different energy thresholds of the various telescope types, the LSTs, although small in number, have a big impact on the reconstruction of shower directions. However, by applying rather soft cuts on the reconstructed shower energies, the number of showers detected with a high multiplicity of telescopes is increased, and thus the direction reconstruction can improve significantly. Detailed results of these studies will be presented elsewhere.

\section*{Acknowledgement} We gratefully acknowledge support from the agencies and organizations 
listed under Funding Agencies at this website: http://www.cta-observatory.org/. SE, AZ, JV and CvE acknowledge financial support from the German BMBF Verbundforschung (Grant ID 05A14WE2).


\begin{thebibliography}{99}
\bibitem{Actis2011}
{Actis}, M., {Agnetta}, G., {Aharonian}, F., {et~al.}
\newblock {Design concepts for the Cherenkov Telescope Array CTA: an advanced
	facility for ground-based high-energy gamma-ray astronomy}.
\newblock \emph{Experimental Astronomy}, 32:\penalty0 193--316, Dec. 2011.
\newblock \doi{10.1007/s10686-011-9247-0}.

\bibitem{root}
\href{https://root.cern.ch/drupal/}{https://root.cern.ch/drupal/}

\bibitem{Bernloehr2013}
{Bernl{\"o}hr}, K., {Barnacka}, A., {Becherini}, Y., {et~al.}
\newblock {Monte Carlo design studies for the Cherenkov Telescope Array}.
\newblock \emph{Astroparticle Physics}, 43:\penalty0 171--188, Mar. 2013.
\newblock \doi{10.1016/j.astropartphys.2012.10.002}.

\bibitem{Naurois2009}
{{de Naurois}}, M. \& {Rolland}, L.
\newblock {A High Performance Likelihood Reconstruction of {$\gamma$}-Rays for
	Imaging Atmospheric Cherenkov Telescopes}.
\newblock \emph{Astroparticle Physics}, 32:\penalty0 231--252, {Dez.} 2009.
\newblock \doi{10.1016/j.astropartphys.2009.09.001}.

\end{thebibliography}
\end{document}